\documentclass[reprint,pra,shownopacs,twocolumn,superscriptaddress]{revtex4-1}

\usepackage{graphicx}
\usepackage{dcolumn}
\usepackage{bm}
\usepackage{mathrsfs}
\usepackage{color}
\usepackage{epstopdf}
\usepackage{booktabs}
\usepackage{amsmath}

\newcommand{\ba}{\begin{array}}
\newcommand{\ea}{\end{array}}
\newcommand{\be}{\begin{equation}}
\newcommand{\ee}{\end{equation}}
\newcommand{\bea}{\begin{eqnarray}}
\newcommand{\eea}{\end{eqnarray}}

\newcommand{\bfD}{{\bf D}}

\newcommand{\bfe}{{\bf e}}
\newcommand{\bfn}{{\bf n}}

\newcommand{\ri}{r_{\rm i}}

\newcommand{\tU}{{\widetilde U}}

\begin{document}

\title{Active rejection-enhancement of spectrally adaptive liquid crystal geometric phase vortex coronagraphs}

\author{Nina Kravets}
\affiliation{Universit{\'e} de Bordeaux, CNRS, Laboratoire Ondes et Mati{\`e}re d'Aquitaine, F-33400 Talence France}
\author{Urban Mur}
\affiliation{Faculty of Mathematics and Physics, University of Ljubljana, Jadranska 19, 1000 Ljubljana, Slovenia}
\author{Miha Ravnik}
\affiliation{Faculty of Mathematics and Physics, University of Ljubljana, Jadranska 19, 1000 Ljubljana, Slovenia}
\affiliation{Jo{\v z}ef Stefan Institute, Jamova 39,1000 Ljubljana, Slovenia}
\author{Slobodan {\v Z}umer}
\affiliation{Jo{\v z}ef Stefan Institute, Jamova 39,1000 Ljubljana, Slovenia}
\affiliation{Faculty of Mathematics and Physics, University of Ljubljana, Jadranska 19, 1000 Ljubljana, Slovenia}
\author{Etienne Brasselet}
\email{etienne.brasselet@u-bordeaux.fr}
\affiliation{Universit{\'e} de Bordeaux, CNRS, Laboratoire Ondes et Mati{\`e}re d'Aquitaine, F-33400 Talence France}

\begin{abstract}
Geometric phase optical elements made of space-variant anisotropic media customarily find their optimal operating conditions when the half-wave retardance condition is fulfilled, which allows imparting polarization-dependent changes to an incident wavefront. In practice, intrinsic limitations of man-made manufacturing process or the finite spectrum of the light source lead to a deviation from the ideal behavior. This implies the implementation of strategies to compensate for the associated efficiency losses. Here we report on how the intrinsic tunable features of self-engineered liquid crystal topological defects can be used to enhance the rejection capabilities of spectrally adaptive vector vortex coronagraphs. We also discuss the extent of which current models enable to design efficient devices.
\end{abstract}

\maketitle

Electrical and optical properties of liquid crystals make them attractive material for numerous photonics applications requiring remote changes of the phase or the polarization state of light. This is usually achieved by electrically driven modification of orientational state of the material along the light propagation direction. The transverse spatial orientation state of material brings additional degree of freedom for phase modulation. This is the case for liquid crystal slabs whose orientational state (defined locally by the average molecular orientation to which we assign a unit vector $\bfn$ called director) imparts a half-wave birefringent retardation to an incident light field along its propagation direction and space-variant phase profile of a geometric nature in the transverse plane. Their complex amplitude transfer function is given as $t = \exp(\pm 2 i \psi)$, where $\pm$ refers to the handedness of the incident circular polarization state and $\psi$ is its effective in-plane optical axis orientation angle. Such geometric phase optical elements were anticipated in the late 1990s \cite{bhandari_pr_1997} and experimentally realized a few years after using space-variant solid-state subwavelength gratings \cite{bomzon_ol_2002, biener_ol_2002} while the advent of their liquid crystal counterparts appeared only a few years later \cite{honma_jjap_2005, marrucci_prl_2006}.

In the present work we focus on liquid crystal geometric phase optical vortex generators ideally associated with $t = \exp(\pm 2im\phi)$, where $m$ is an interger or half-integer and $\phi$ is the polar angle in the $(x,y)$ plane of the optical element, whose technological maturity and field of use have continued to grow since their first realization in 2006 \cite{marrucci_prl_2006}. To date there is a trade-off between, on the one hand, the tunability of the operating wavelength in order to satisfy the half-wave plate condition, which is achieved either by thermal \cite{karimi_apl_2009} or electrical \cite{piccirillo_apl_2010} means and, on the other hand, the spatial resolution of the structural singularity for the director orientation that can reach sub-micrometer size \cite{tabirian_ieee_2015}. Since most applications involve optical beams with cross sections in the millimeter range or larger, they do not suffer from such a compromise. However, this is a priori no longer true when the geometric phase vortex mask is required in the focal plane of an optical system, as is the case for vector optical vortex coronagraphy.

Optical coronagraphy is a high contrast imaging technique originally developed more than 80 years ago to create artificial total eclipses of the Sun \cite{lyot_mnras_1939}. The working principle of the original apparatus is to occult the central part of the Airy diffraction pattern in the Fourier plane of a telescope in order to strongly reduce the amount of on-axis stellar light reaching the observer. It was not until nearly sixty years later that the (binary) manipulation of the phase distribution of the Airy spot rather than its occultation was considered \cite{roddier_pasp_1997}. Further developments led to the advent of optical vortex coronagraphy where a vortex phase mask with integer values of $m$ is centered on the Airy pattern of the on-axis light source to be rejected. Scalar and vector coronagraphy have been proposed simultaneously in Refs.~\citenum{foo_ol_2005} and \citenum{mawet_apj_2005}. Their respective labels refer to the physical origin of the phase changes operated by the vortex phase mask: dynamic (scalar case) or geometric (vector case). The superior chromatic performances of the geometric phase option makes vector vortex coronagrahy more attractive. Nowadays, more than one decade after first laboratory \cite{mawet_oe_2009} and astronomical \cite{mawet_apj_2010, serabyn_nature_2010} demonstrations, {\it spectrally static} geometric phase vortex masks equip several ground based astronomical observations facilities \cite{absil_spie_2016}. Still, it is important to note that spectroscopic imaging of extrasolar planets to learn about their atmospheric composition involves developing broadband vortex masks. This inherently comes with polarization leakage problems and requires demanding technological improvements to mitigate the effects, not only for the vortex mask itself but also for the additional polarization optics involved \cite{serabyn_josab_2019, doelman_pasp_2020}.

Instead, another approach would be to consider {\it spectrally adaptive} geometric phase vortex masks that do not require the use of additional polarization optics and to use them in a narrowband regime while the operating wavelength is scanning the desired spectral bandwidth. However, spectrally adaptive liquid crystal geometric phase vortex coronagraphs reported so far remain hampered by central disorientation region trade-off \cite{aleksanyan_ol_2016, aleksanyan_prl_2017, piccirillo_mclc_2019}. Here we report on how the rejection capabilities of the latter devices can be enhanced.

Noteworthy, there are two distinct kinds of central disorientation in liquid crystal geometric phase vortex masks: (i) the departure from the space-variant pattern $\psi = m\phi$ (to an unimportant constant) and (ii) the departure from the half-wave birefringent retardance, which could mix in practice. The few previous attempts can be classified as being mainly either on type (i) \cite{piccirillo_mclc_2019} or type (ii) \cite{aleksanyan_ol_2016, aleksanyan_prl_2017}. In the first case, it is difficult to consider post-reprogramming of the fabrication-limited patterned anchoring layers that provide liquid crystal alignment. Nevertheless, the placement of an opaque disk covering the troublesome region is a rough solution applicable regardless of the nature of the vortex mask \cite{mawet_apj_2010}. In the second case, which refers to the use of spontaneously formed liquid crystal topological defects under the action of external fields, an additional backup option consists to place the vortex masks between crossed circular polarizers at the expense of throughput losses of at least 50\% for unpolarized light observations and additional polarization optics chromatic issues.

\begin{figure}[t!]
\centering
\includegraphics[width=1\columnwidth]{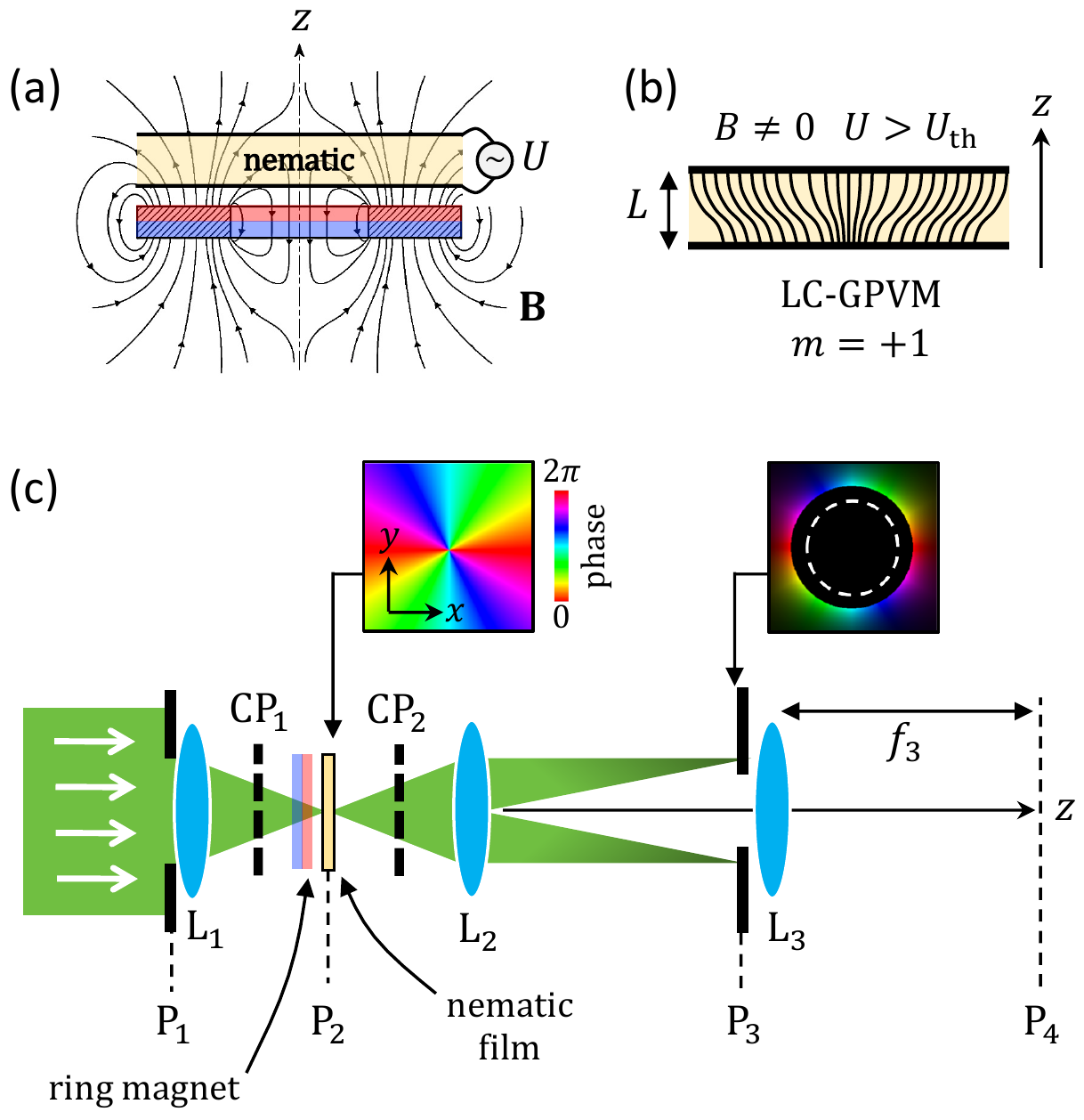}
\caption{
(a) Illustrations of the magneto-electric configuration to generate on-demand macroscopic umbilic defect in a nematic liquid crystal film used as a spectrally adaptive liquid crystal geometric phase vortex mask (LC-GPVM) with $m=+1$. The Nickel-plated Neodymium (grade N50) ring magnet (not on scale) has internal and external radii $R_{\rm int}=2~$mm and $R_{\rm ext}=6~$mm, respectively, and height $H=6~$mm. Its magnetization is directed along the $z$ axis and is associated with a pull force of $\sim 32$~N (manufacturer datasheet). The magnet is placed at a distance $\sim 2$~mm from the input facet of the nematic slab. (b) Side view sketch of the lines of the ideally axisymmetric director field under the combined action of the magnetic and electric field. (c) Sketch of the coronagraphic experimental setup. P$_i$: planes $i=(1,2,3,4)$ of interest; L$_i$: lens $i=(1,2,3,4)$ with focal length f$_i$; CP$_{1,2}$: circular polarizers. Parameters: the radius of the circular aperture at P$_1$ is $R_1=1$~mm, the radius of the circular aperture at P$_3$ is $R_2=0.75 R_1$, the focal length are $f_1=f_2=200$~mm and $f_3=400$~mm. All the lenses are achromatic doublets. See text for detailed description. Inset at P$_2$: phase map of the optical vortex mask. Inset at P$_3$: calculated intensity and phase, where the luminance refers to the intensity and the colormap refers to the phase. The dashed circle refers to the aperture with radius $R_2$.
}
\label{fig:setup}
\end{figure}

In order to get rid of obstruction and polarization filtering backup strategies for type (ii) vortex masks, one of us suggested a few years ago that adaptive optimization of the size of the disorientation region is an open option when using liquid crystal topological defects called umbilics \cite{brasselet_prl_2018}, which is quantitatively explored experimentally in the present work. Also, we discuss the capabilities of available analytical model and full numerical simulation to describe adaptive downsizing of the core of umbilics towards optimal design. Umbilics are nonsingular topological defects associated with $m = \pm 1$ unveiled 40 years ago by Rapini \cite{rapini_jp_1973} in nematic liquid crystal having negative dielectric anisotropy ($\epsilon_{\rm a}<0$) and sandwiched between two parallel substrates providing uniform perpendicular orientational boundary conditions for the director ($\bfn (x,y,z=0) = \bfn (x,y,z=L) = \bfe_z$ where $L$ is the cell thickness and $\bfe_z$ the unit vector along the $z$ axis). These defects spontaneously appear when applying a quasistatic voltage between the two facets of the nematic slab that exceeds the Fr{\'e}edericksz threshold value $U_{\rm th} = \pi\sqrt{K_3/(\epsilon_0|\epsilon_{\rm a}|)}$ where $K_3$ is the bend elastic constant of the nematic and $\epsilon_0$ is the vacuum dielectric permittivity \cite{rapini_jp_1973}. In our experiments we used a $20~\mu$m-thick sample prepared with the dual frequency nematic mixture 1859A (from Military University of Technology, Warsaw, Poland) and umbilics are obtained following the magnetic-electric approach proposed in \cite{brasselet_prl_2018}, see Figs.~1(a) and 1(b). The combined action of a static magnetic field from a ring magnet with a quasistatic electric field (square waveform at 200~kHz frequency) enables robust self-engineering of geometric phase vortex masks with $m=1$ above $U_{\rm th} = 2.85$~V.

\begin{figure*}[t!]
\centering
\includegraphics[width=2\columnwidth]{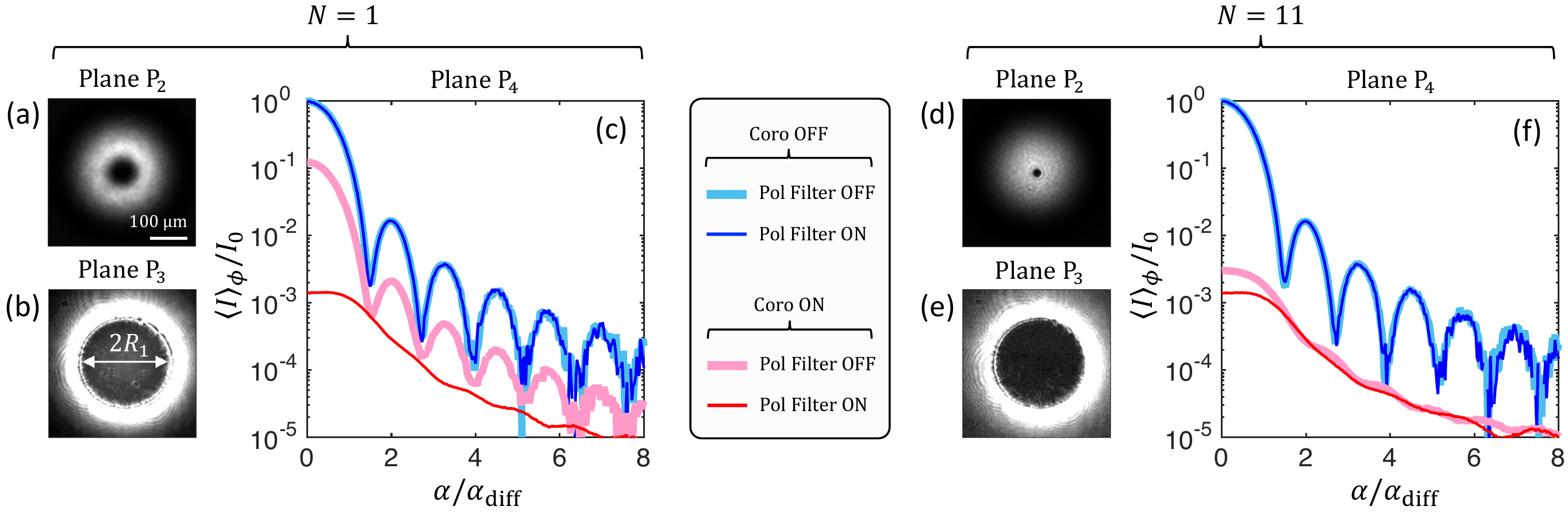}
\caption{
Left and right parts: $\Delta_\infty=\pi$ ($N=1$) and $\Delta_\infty=11\pi$ ($N=11$), respectively. (a,d) Airy spot in P$_2$ observed between crossed circular polarizers. (b,e) Ring of fire without polarization filterting in P$_3$. (c,f) Azimuth-average angular intensity profile of the observed point-like source in P$_4$ with (`coro ON') and without (`coro OFF') activating the coronagraphic mask (i.e., by centering or laterally shifting the center of the umbilic with the Airy spot in P$_2$) and with (`Pol Filter ON') and without (`Pol Filter OFF') polarization filtering, where $\alpha$ is the polar angle in P$_4$ and $\alpha_{\rm diff}=0.61\lambda/R_1$ is the is the diffraction limit angle of the telescope. Parameters: wavelength $\lambda=633$~nm at temperature $T=19^\circ$C and $U=3.09~$V for $N=1$, and at $T=45^\circ$C and $U=14.80~$V for $N=11$. Importantly, we keep $U<18$~V in order to prevent from electrical destabilization related to the imperfections of the liquid crystal sample, while the temperature is adapted to preserve the half-wave retardance condition $\Delta_\infty=N\pi$ with $N$ odd integer. The corresponding pairs of images [(a) and (d), (b) and(e)] are respectively recorded with equal exposure time and input light power for direct visual comparison while (b) and(e) are overexposed to emphasize the degraded ring of fire for the lowest value of $N$.
}
\label{fig:coro_exp}
\end{figure*}

As shown in Fig.~1(c), the mask is placed in the Fourier plane (P$_2$) of a lens (L$_1$, focal length $f_1$) illuminated by a collimated expanded laser beam from a supercontinuum source that can be spectrally filtered on-demand using a set of bandpass interferential filters. The input pupil plane P$_1$ is located right before L$_1$ where a metallic circular aperture with radius $R_1$ is placed. A second lens (L$_2$, focal length $f_2$) placed at a distance $f_2$ from P$_2$ produces the image of the input pupil in the plane P$_3$ located at a distance $f_2(1+f_2/f_1)$ from L$_2$, where a metallic circular aperture with radius $R_2$ is placed. Ideally, full rejection of on-axis incident light is achieved when the Airy spot is centered on the vortex phase mask placed in P$_2$ provided that $R_2<(f_2/f_1)R_1$ \cite{mawet_apj_2005}, as illustrated by the insets of Fig.~1(c). The emulated stellar imaging is made using a third lens (L$_3$, focal length $f_3$) by placing a camera in the observation plane P$_4$ located at a distance $f_3$ from L$_3$. To date, optical vortex coronagraphy using umbilics as vortex masks have been made by adjusting the half-wave retardance criterion in the zeroth-order condition for monochromatic laser sources \cite{aleksanyan_ol_2016, aleksanyan_prl_2017}. This corresponds to birefringent phase retardation $\Delta$ set at the asymptotic value $\Delta_\infty = N\pi$, with $N=1$, sufficiently far from the center of the umbilic defect. The axisymmetry of the umbilics with $m=1$ imposes an optical retardance that vanishes as the distance $r$ from the defect center tends to zero, causing the optical vortex mask to deviate from the ideal condition of uniform half-wave retardance. Its detrimental effect is managed by placing the liquid crystal mask between two crossed circular polarizers, which leads to $\sim 10^3$ peak-to-peak intensity reduction for the azimuth average coronagraphic images in P$_4$ as reported in Ref.~\citenum{aleksanyan_ol_2016, aleksanyan_prl_2017}.

Here our sample provides similar performances for $N=1$ (reached at $U \simeq 1.1 U_{\rm th}$) as in our previous works when polarization filtering is activated, see thin curves in Fig.~2(c). In other words, polarization filtering is necessary regardless of the way used (electric, photo-electric or magneto-electric) to create the umbilics-based mask. Once polarization filtering is removed, the coronagraphic performance drastically drops, see thick curves in Fig.~2(c). This visually can be grasped from imperfect `ring of fire' in P$_3$ shown in Fig.~2(b) where one can see that intensity leaks inside the re-imaged area of the input pupil plane ($r<R_1$). This is related to the fact that the size of the nonuniform central part of the liquid crystal vortex mask is not negligible compared to that of the Airy spot, see Fig.~2(a). Here we show that increasing the voltage sufficiently far from $U_{\rm th}$ we can shrink the nonuniform central part of the mask which makes the polarization filtering not needed. The results are shown in Fig.~2(d--f) for $U \simeq 5.2 U_{\rm th}$, which corresponds to the high-order half-wave condition with $N=11$.

\begin{figure}[t!]
\centering
\includegraphics[width=1\columnwidth]{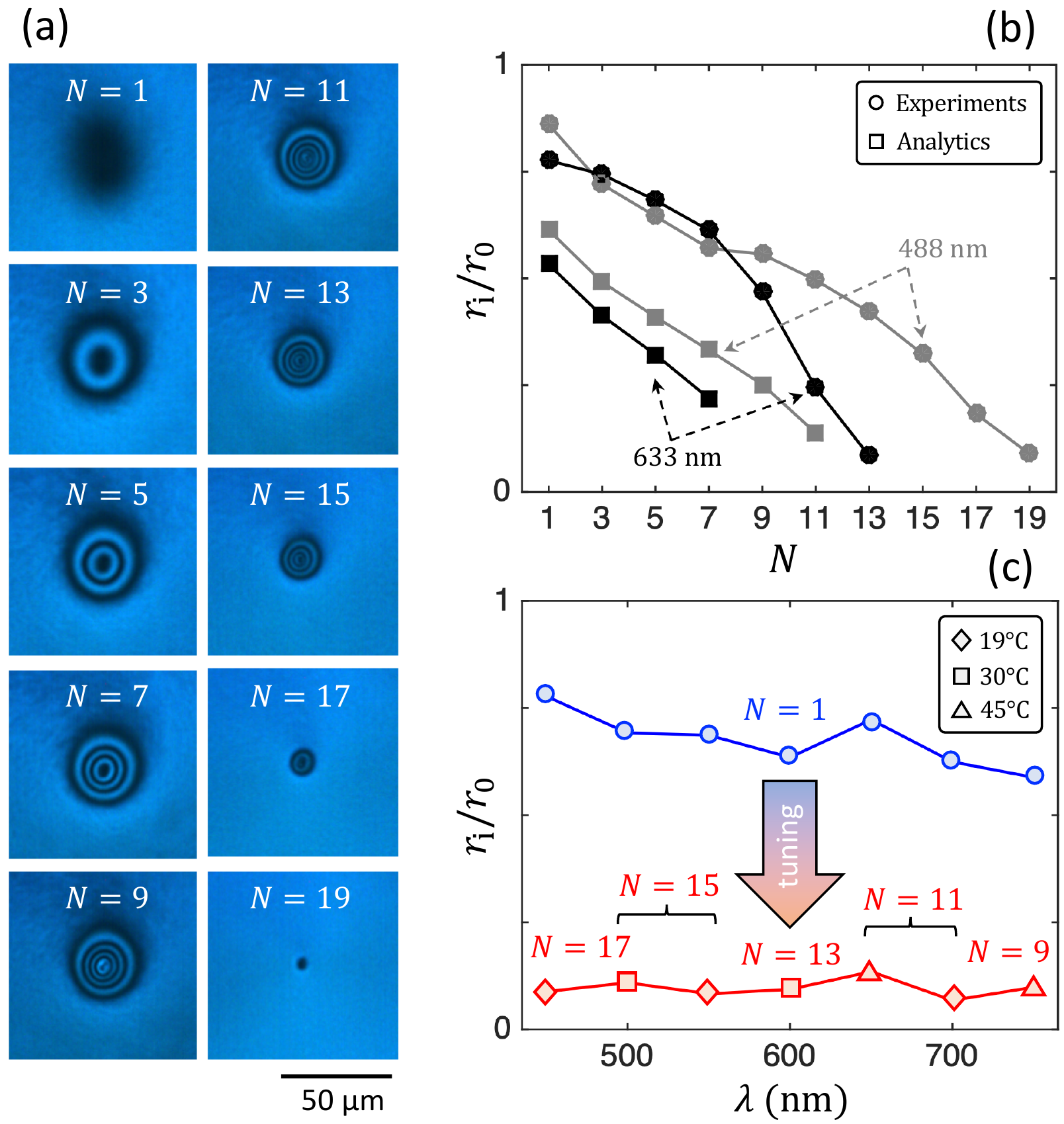}
\caption{
(a) Images of the central part of the vortex mask observed between crossed circular polarizers using incoherent white light source spectrally filtered at 488~nm wavelength and a $20\times$ microscope objective with numerical aperture ${\rm NA} = 0.3$, at $T=19^\circ$C. Ten orders are observed for the half-wave behavior (hence ten odd values of $N$ from 1 to 19) as the applied electric field is increased at room temperature. (b) Measured dependence of $\ri$ versus $N$ at 488~nm and 633~nm wavelengths (circle markers) and the corresponding analytical expectations from Rapini's model (square markers). (c) Electro-thermal adaptive optimization of the spectral performance in the visible domain.
}
\label{fig:ri_exp}
\end{figure}

Specifically, the relevant quantity allowing to assess the downsizing of the nonuniform central part of the liquid crystal vortex mask is of light-matter nature. In fact, birefringent phase retardation modulation takes place along the radial coordinate in the vicinity of the center of the umbilic, as illustrated in Fig.~3(a) where the vortex mask is imaged between crossed circular polarizers as $N$ is increased by electrical means at fixed temperature. Qualitatively, the vortex mask can be considered as being uniform at a given half-wave retardance order $N$ at distance $r$ larger than the radius of the last dark ring (or dark area for $N=1$ and $N \gg 1$). Quantitatively, this is retrieved from azimuth-average radial intensity profiles by defining the radius $\ri$ of inhomogeneous retardance as the largest value of $r$ satisfying $\langle I(r) \rangle_\phi = 0.95 \max_r[\langle I(r) \rangle_\phi]$. The results are shown in Fig.~3(b) at 488~nm and 633~nm wavelengths, see circle markers. The maximal odd values of $N$ that can be reached experimentally depends on the wavelength: $N_{\max}=19$ for $\lambda=488$~nm and $N_{\max}=13$ for $\lambda=633$~nm. These values are consistent with the expected asymptotic birefringent phase retardation $\Delta_\infty = 2\pi dnL/\lambda \simeq 18.8 \pi$ ($\lambda=488$~nm) and $14.5 \pi$ ($\lambda=633$~nm) assuming that the director lies in the plane of the mask at all $z$ when $\tU \gg 1$, where $\tU = U/U_{\rm th}$, and taking $dn = 0.2287$ (at $\lambda=589$~nm and $T = 20^\circ$C, manufacturer datasheet). Such a trend can be grasped from Rapini's analytical model from which the function $\Delta(r)$ can be calculated, hence $\ri = \max_r \{ \sin^2[\Delta(r)/2] = 0.95\}$, see Refs.~\citenum{rapini_jp_1973, brasselet_prl_2012} for calculation details. Analytical predictions are shown as square markers in Fig.~3(b), which offers a qualitative rather than a quantitative match with our observations. This can be understood recalling that Rapini's model assumes monomodal longitudinal profile for the director tilt angle $\theta$ with respect to the $z$ axis, namely, $\theta(r,z) \propto \sin(\pi z/L)$, while experimental data support a saturation phenomenon, $\theta \to \pi/2$ everywhere, as the magnitude of the electrical torque exerted on the liquid crystal increases.

On-demand optimization ensuring high-order half-wave retardance condition and significant coronagraphic rejection without need of polarization filtering is therefore possible for any operating wavelength. This is illustrated in Fig.~3(c) where the applied electric field and temperature are used as tuning parameters. We reached $\ri / r_0 = (9.6 \pm 2.1)\times10^{-2}$, where $r_0 = 0.61\lambda f_1/R_1$ is Airy disc radius, over the wavelength range $450~{\rm nm} < \lambda < 750~{\rm nm}$ that covers almost all the visible domain.

Noting that Rapini's model is not suitable quantitative for the present purpose, we argue that theoretical design of optimized adaptive masks requires more elaborated treatment (Rapini's model is only valid for $\theta^2 \ll 1$). This can be achieved by relying on Landau-de Gennes free energy minimization full-numerical approach, as reviewed in Ref.~\citenum{ravnik_lc_2009}. In contrast to Rapini's model, a numerical approach allows to treat exactly the field-matter dielectric coupling without resorting on discarding transverse contributions to the constitutive Maxwell's equation $\nabla \cdot \bfD=0$. Specifically, we jointly solve the full Landau-de Gennes free energy minimization and the generalized Laplace equation $\partial_i \epsilon_{ij} \partial_j V=0$ for the electric potential $V$ in the liquid crystal slab, associated with the boundary conditions $V(x,y,z=0)=0$ and $V(x,y,z=L)=U$. Dielectric permittivity tensor $\epsilon_{ij}$ is directly related to the nematic order parameter tensor $Q_{ij}$ as $\epsilon_{ij}=\epsilon_{\rm iso} \delta_{ij} +\frac{2}{3}\epsilon_a^{\rm mol} Q_{ij}$, where $\epsilon_{\rm iso}$ is isotropic dielectric permittivity, $\epsilon_a^{\rm mol}$ is molecular dielectric anisotropy, $i,j=(x,y,z)$, and summation over double indices is assumed. Also, from the material point of view, the numerical approach allows accounting for possible changes of nematic order in presence of large orientational gradients and handling orientational saturation effects.

\begin{figure}[t!]
\centering
\includegraphics[width=1\columnwidth]{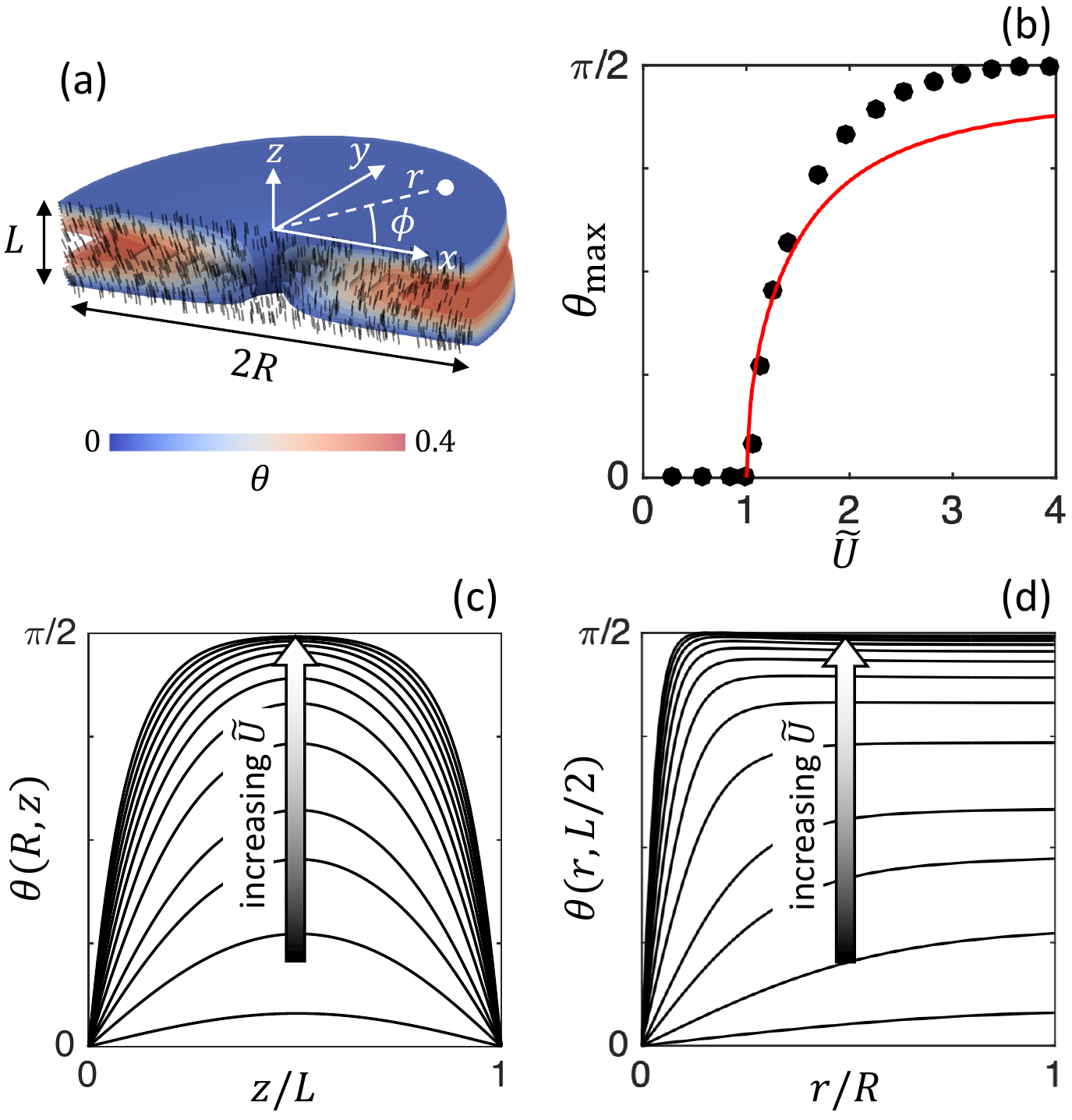}
\caption{
(a) Typical three-dimensional director field profile of a numerically simulated umbilic defect using tensorial order parameter Landau-de Gennes formalism coupled to generalized Laplace equation, at $\tU=1.12$. Black segments refers to the orientation of the director $\bfn$ and color levels refer to the value of the director tilt angle $\theta$. (b) Maximal director reorientation angle versus the reduced applied voltage. Black markers: simulations. Red curve: Rapini's model. (c,d) Saturation of the longitudinal and transverse, respectively, director reorientation profiles as $\tU$ increases according to the 13 values above threshold reported in panel (a): $\tU = 1.05$, 1.12, 1.26, 1.40, 1.68, 1.96, 2.24, 2.52, 2.81, 3.09, 3.37, 3.65 and 3.93.
}
\label{fig:numerics}
\end{figure}

We choose a parallelepipedic simulation box with square cross-section and assume fixed orientational boundary conditions at the edges of the box in all directions, where we imposed $\bfn = \bfe_z$, respectively. The free energy minimum is found by solving the Euler-Lagrange equations together with the generalized Laplace's equation, which eventually gives the director field. A finite-difference relaxation method is used in a cylindrical domain within a computational box with the size of $750\times750\times100$ voxels, with a spatial resolution of 10~nm in all directions. We note that computational resources prevent to address the behavior of a sample with the same dimensions as those used in the experiments; however, the demonstrated results to quantitatively describe the adaptive liquid crystal geometric phase vortex mask allows to highlight the need of going beyond Rapini's model towards designing optimal elements. During the relaxation, the director field on all lattice sites is updated in each time step until the steady state is achieved, usually after a few $10^5$ relaxation steps. This gives reliable results inside a cylindrical domain with radius $R=2.5~\mu$m for a slab of thickness $L = 0.94~\mu$m, see Fig.~4(a). 
Indeed, 3 layers of voxels are used on the top and bottom parts of the computational box in order to ensure in the numerical formulation the strong boundary conditions while 125 voxels in the radial direction are found necessary to screen the effects of finite extension of the nematic slab in the $(x,y)$ plane.

The results are shown in Fig.~4, using single elastic constant $K = 12.8$~pN (neglecting elastic anisotropy), $\epsilon_\perp=4.33$ and $\epsilon_\parallel=2.25$. In Fig.~4(b) the numerical predictions are compared to Rapini's model (red curve) that gives $ \theta_{\max} = \{2(\tU^2 - 1)/[1 - (\epsilon_a/\epsilon_\parallel)\tU^2]\}^{1/2}$ accounting for the single elastic constant approximation, see Eq.~(13) in Ref.~\citenum{rapini_jp_1973}. In Fig.~4(c) the simulations clearly exhibit the appearance of high-order odd Fourier components for the longitudinal profile of the director as $\tU$ increases, namely, $\theta(r,z) \propto \sum_{n\,{\rm odd}} \theta_n \sin(n \pi z/L)$ with $n>1$. Eventually, Fig.~4(d) displays the radial profile of the director reorientation in the mid-plane of the liquid crystal slab, which we found to moderately depart from the universal profile provided by Rapini's model as $\tU$ increases.

Summarizing, active enhancement of coronagraphic rejection of adaptive liquid crystal geometric phase vortex masks owing to the combined action of applied electric field and temperature has been demonstrated. The approach consists of reducing the size of the non-ideal central region of the vortex mask, thus getting rid of the polarization filtering otherwise required. The strategy is therefore equally applicable to all wavelengths and do not require the use of additional optical components such as occulting mask \cite{mawet_apj_2010}, achromatic polarization optics \cite{serabyn_josab_2019} or additional grating \cite{doelman_pasp_2020}. As such, present work contributes to the emergence of adaptive optical vortex coronagraphy which is so far restricted to given a spectral operating condition.

\vspace{2mm}
N.K. and E.B. thank the support of Conseil R{\'e}gional de Nouvelle Aquitaine (project HELIXOPTICS). U.M, M.R. and S.Z. acknowledge funding from Slovenian research agency ARRS grants P1-0099, N1-0195 and J1-2462, and EU ERC AdG LOGOS.


\begin{thebibliography}{}

\bibitem{bhandari_pr_1997}
R.~Bhandari,{\it Polarization of light and topological phases,} Phys. Rep.
  \textbf{281}, 1--64 (1997).

\bibitem{bomzon_ol_2002}
Z.~Bomzon, G.~Biener, V.~Kleiner, and E.~Hasman,{\it Space-variant
  pancharatnam-berry phase optical elements with computer-generated
  subwavelength gratings,} Opt. Lett. \textbf{27}, 1141 (2002).

\bibitem{biener_ol_2002}
G.~Biener, A.~Niv, V.~Kleiner, and E.~Hasman,{\it Formation of helical
  beams by use of pancharatnam-berry phase optical elements,} Opt. Lett.
  \textbf{27}, 1875 (2002).

\bibitem{honma_jjap_2005}
M.~Honma and T.~Nose,{\it Liquid-crystal fresnel zone plate fabricated by
  microrubbing,} Jpn. J. Appl. Phys. \textbf{44}, 287 (2005).

\bibitem{marrucci_prl_2006}
L.~Marrucci, C.~Manzo, and D.~Paparo,{\it Optical spin-to-orbital angular
  momentum conversion in inhomogeneous anisotropic media,} Phys. Rev. Lett.
  \textbf{96}, 163905 (2006).

\bibitem{karimi_apl_2009}
E.~Karimi, B.~Piccirillo, E.~Nagali, L.~Marrucci, and E.~Santamato,
 {\it Efficient generation and sorting of orbital angular momentum
  eigenmodes of light by thermally tuned q-plates,} Appl. Phys. Lett.
  \textbf{94}, 231124 (2009).

\bibitem{piccirillo_apl_2010}
B.~Piccirillo, V.~D'Ambrosio, S.~Slussarenko, L.~Marrucci, and E.~Santamato,
 {\it Photon spin-to-orbital angular momentum conversion via an
  electrically tunable q-plate,} Appl. Phys. Lett. \textbf{97}, 241104 (2010).

\bibitem{tabirian_ieee_2015}
N.~Tabirian, H.~Xianyu, and E.~Serabyn,{\it Liquid crystal polymer vector
  vortex waveplates with sub-micrometer singularity,} Proceedings of Aerospace
  Conference IEEE pp. 1--10 (2015). DOI: 10.1109/AERO.2015.7119168.

\bibitem{lyot_mnras_1939}
B.~Lyot,{\it A study of the solar corona and prominences without
  eclipses,} Mon. Not. R. Astron. Soc. \textbf{99}, 580--594 (1939).

\bibitem{roddier_pasp_1997}
F.~Roddier and C.~Roddier,{\it Stellar coronograph with phase mask,} Publ.
  Astron. Soc. Pac. \textbf{109}, 815 (1997).

\bibitem{foo_ol_2005}
G.~Foo, D.~M. Palacios, and G.~A. Swartzlander,{\it Optical vortex
  coronagraph,} Opt. Lett. \textbf{30}, 3308--3310 (2005).

\bibitem{mawet_apj_2005}
D.~Mawet, P.~Riaud, O.~Absil, and J.~Surdej,{\it Annular groove phase mask
  coronagraph,} Astrophys. J. \textbf{633}, 1191--1200 (2005).

\bibitem{mawet_oe_2009}
D.~Mawet, E.~Serabyn, K.~Liewer, C.~Hanot, S.~McEldowney, D.~Shemo, and N.~E.
  O'Brien,{\it Optical vectorial vortex coronagraphs using liquid crystal
  polymers: theory, manufacturing and laboratory demonstration,} Opt. Express
  \textbf{17}, 1902--1918 (2009).

\bibitem{mawet_apj_2010}
D.~Mawet, E.~Serabyn, K.~Liewer, K.~Burruss, J.~Hickey, and D.~Shemo,
 {\it The vector vortex coronagraph: Laboratory results and first light at
  palomar observatory,} Astrophys. J. \textbf{709}, 53--57 (2010).

\bibitem{serabyn_nature_2010}
E.~Serabyn, D.~Mawet, and R.~Burruss,{\it An image of an exoplanet
  separated by two diffraction beamwidths from a star,} Nature \textbf{464},
  1018--1020 (2010).

\bibitem{absil_spie_2016}
O.~Absil, D.~Mawet, M.~Karlsson, B.~Carlomagno, V.~Christiaens, D.~Defr{\`e}re,
  C.~Delacroix, B.~Femen{\'i}a~Castell{\'a}, P.~Forsberg, J.~Girard, C.~A.
  G{\'o}mez~Gonz{\'a}lez, S.~Habraken, P.~M. Hinz, E.~Huby, A.~Jolivet,
  K.~Matthews, J.~Milli, G.~Orban~de Xivry, E.~Pantin, P.~Piron, M.~Reggiani,
  G.~J. Ruane, G.~Serabyn, J.~Surdej, K.~R.~W. Tristram, E.~Vargas~Catal{\'a}n,
  O.~Wertz, and P.~Wizinowich,{\it Three years of harvest with the vector
  vortex coronagraph in the thermal infrared,} Proc. SPIE \textbf{9908}, 99080Q
  (2016).

\bibitem{serabyn_josab_2019}
E.~Serabyn, C.~M. Prada, P.~Chen, and D.~Mawet,{\it Vector vortex
  coronagraphy for exoplanet detection with spatially variant diffractive
  waveplates,} J. Opt. Soc. Am. B \textbf{36}, D13--D19 (2019).

\bibitem{doelman_pasp_2020}
D.~S. Doelman, E.~H. Por, G.~Ruane, M.~J. Escuti, and F.~Snik,
 {\it Minimizing the polarization leakage of geometric-phase coronagraphs
  with multiple grating pattern combinations,} Publ. Astron. Soc. Pac.
  \textbf{132}, 045002 (2020).

\bibitem{aleksanyan_ol_2016}
A.~Aleksanyan and E.~Brasselet,{\it Vortex coronagraphy from
  self-engineered liquid crystal spin-orbit masks,} Opt. Lett. \textbf{41},
  5234--5237 (2016).

\bibitem{aleksanyan_prl_2017}
A.~Aleksanyan, N.~Kravets, and E.~Brasselet,{\it Multiple-star system
  adaptive vortex coronagraphy using a liquid crystal light valve,} Phys. Rev.
  Lett. \textbf{118}, 203902 (2017).

\bibitem{piccirillo_mclc_2019}
B.~Piccirillo, E.~Piedipalumbo, L.~Marrucci, and E.~Santamato,
 {\it Electrically tunable vector vortex coronagraphs based on
  liquid-crystal geometric phase waveplates,} Mol. Cryst. Liq. Cryst.
  \textbf{684}, 15--23 (2019).

\bibitem{brasselet_prl_2018}
E.~Brasselet,{\it Tunable high-resolution macroscopic self-engineered
  geometric phase optical elements,} Phys. Rev. Lett. \textbf{121}, 033901
  (2018).

\bibitem{rapini_jp_1973}
A.~Rapini,{\it Umbilics : Static properties and shear-induced
  displacements,} J. Phys. \textbf{34}, 629--633 (1973).

\bibitem{brasselet_prl_2012}
E.~Brasselet,{\it Tunable optical vortex arrays from a single nematic
  topological defect,} Phys. Rev. Lett. \textbf{108}, 087801 (2012).

\bibitem{ravnik_lc_2009}
M.~Ravnik and S.~{\v{Z}}umer,{\it Landau--de gennes modelling of nematic
  liquid crystal colloids,} Liq. Cryst. \textbf{36}, 1201--1214 (2009).

\end{thebibliography}

\end{document}